\documentclass[aps,pre,twocolumn,amsmath,amssymb,superscriptaddress,floatfix]{revtex4-2}

\usepackage{graphicx}
\usepackage{bm}
\usepackage{mathtools}
\IfFileExists{microtype.sty}{\usepackage{microtype}}{}
\usepackage{amsthm}

\newtheorem{proposition}{Proposition}

\newcommand{\E}{\mathbb{E}}
\newcommand{\Var}{\mathrm{Var}}
\newcommand{\Prob}{\mathbb{P}}

\begin{document}

\title{Analytically tractable model of synaptic crowding explains emergent small-world structure and network dynamics}

\author{Makoto Fukushima}
\email{mak@toki.waseda.jp}
\affiliation{Department of Physics, Waseda University, Tokyo, Japan}
\altaffiliation{Current address: Honda Research Institute Japan, Wako, Saitama, Japan}

\date{\today}

\begin{abstract}
Neural circuits must balance local connectivity constraints against the need for global integration. Here we introduce a minimal wiring rule motivated by synaptic crowding: as a neuron accumulates incoming connections, each additional synapse becomes progressively harder to form. This single-parameter model admits an exact finite-size solution for the induced in-degree distribution and yields simple scaling laws: mean connectivity grows only logarithmically with network size while variance remains bounded---consistent with homeostatic regulation of synaptic density. When candidates are encountered in order of spatial proximity, the crowding rule produces a broad, approximately power-law distribution of connection lengths without prescribing any explicit distance-dependent wiring law; combined with shortcut rewiring, this yields networks with small-world characteristics. We further show that the induced degree statistics largely determine attractor basin boundaries in threshold network dynamics, while local clustering primarily modulates the prevalence of long-lived non-absorbing outcomes near these boundaries. The model provides testable predictions linking local developmental constraints to macroscopic network organization and dynamics.
\end{abstract}


\maketitle

\section{Introduction}

Local constraints during synaptogenesis can imprint macroscopic signatures on both network topology and dynamics. Here we show that an elementary synaptic-crowding penalty on incoming edges yields (i) an exactly solvable family of sparse directed graphs with closed-form finite-$N$ degree statistics, and (ii) when combined with distance-ordered candidate encounters, an emergent small-world wiring kernel and multi-scale distribution of connection lengths---without prescribing a distance-dependent connection probability. We further show that the induced degree statistics largely determine basin boundaries in synchronous threshold dynamics, whereas local clustering mainly controls the probability mass in long-lived non-absorbing outcomes near the boundary.

Two ubiquitous constraints motivate the crowding and spatial variants considered here.
First, synaptic resources (molecular machinery, membrane area, metabolic budget) are finite, so adding additional synapses onto a neuron should become progressively harder as synapses accumulate on its dendritic arbor---a ``crowding'' effect that has been discussed in the context of structural plasticity and homeostatic rewiring \cite{Gallinaro2018,Gallinaro2022,Tiddia2024}.
Second, circuits are embedded in physical space: most new synapses form locally, while a minority of long-range projections enable global integration.
At the mesoscale, anatomical and functional brain networks are often reported to exhibit small-world features---high clustering alongside short characteristic path length \cite{WattsStrogatz1998,SpornsZwi2004,BassettBullmore2006}.

Classical small-world generative models either rewire a fraction of local edges (as in the Watts--Strogatz construction) or augment a spatial lattice with long-range edges drawn from an explicit distance-dependent kernel.
A canonical example is Kleinberg's ``navigable'' small-world model, in which long-range links are selected with probability proportional to $d^{-D}$ on a $D$-dimensional lattice \cite{Kleinberg2000}.
Here we take a different approach: we do not prescribe a distance kernel.
Instead, we show that a purely degree-dependent crowding penalty, when combined with the minimal geometric assumption that nearer candidates are encountered earlier during synaptogenesis, induces a Kleinberg-type distance dependence and a broad (approximately log-uniform) distribution of wiring lengths.

From a theoretical standpoint, macroscopic descriptions of randomly connected networks have a long history.
Seminal work by Amari introduced statistical neurodynamics and mean-field state equations for recurrent networks \cite{Amari1971,Amari1972,Amari1974,Amari1977}.
Later developments include associative-memory models \cite{Hopfield1982,AmariMaginu1988} and annealed approximations for random Boolean/threshold networks \cite{DerridaPomeau1986,LuqueSole1997}.
A common simplification is to assume i.i.d.\ couplings or random graphs with externally imposed degree distributions.
In biology, however, degree heterogeneity and multi-scale structure may arise endogenously from developmental mechanisms and geometric constraints.
This motivates random-graph ensembles defined by explicit wiring \emph{rules} rather than by prescribing a distribution.

Several classes of generative models have explored degree-dependent and distance-dependent wiring rules for brain networks.
Hidden-variable (fitness) models \cite{Caldarelli2002,Boguna2003} assign each node an intrinsic parameter that modulates link formation, while the preferential-attachment framework \cite{Barabasi1999} couples link probability to current degree---rewarding rather than penalizing high connectivity.
In the neuroscience literature, generative models combining distance penalties with topological attachment rules have been shown to reproduce key features of empirical connectomes \cite{Vertes2012,Betzel2016,BetzelBassett2017}.
At the circuit level, Peters' rule posits that synapses form in proportion to axo-dendritic overlap \cite{Kalisman2005}, while structural-plasticity models implement homeostatic rules in which synapse creation rates depend on current connectivity \cite{Butz2013}.
Empirically, interareal cortical connection probability has been found to decay exponentially with physical distance \cite{ErcseyRavasz2013}, and the trade-off between wiring cost and topological efficiency is widely considered a key organizing principle \cite{Bullmore2012}.
Our model captures the essential feature common to these frameworks---that the propensity to accept additional synapses decreases with the number already present---while remaining analytically tractable.

Here we study a minimal directed-graph generator that implements a crowding penalty at the level of incoming edges:
when building the in-neighborhood of a target node, each additional accepted edge becomes exponentially less likely.
This single parameter $\alpha$ induces a nontrivial in-degree distribution $P_{\alpha}(k)$ that admits an exact finite-$N$ recursion and a compact generating-function iteration.
We then connect structure to dynamics by analyzing synchronous threshold dynamics on these graphs.
In the thermodynamic limit we obtain a heterogeneous mean-field (HMF) map for the activity fraction.
To capture finite-size effects we introduce an absorbing Markov closure---a binomially driven one-dimensional chain---whose committor function estimates basin (hitting) probabilities.

A key theme is separating what is determined primarily by \emph{degree statistics} from what depends on \emph{geometry and motifs}.
Because the crowding rule depends only on the current accepted in-degree, the distribution of the number of accepted edges is invariant under permutations of the candidate list.
This allows a spatially ordered variant (candidates proposed in increasing distance) and a controlled shortcut-rewiring interpolation that preserve $P_{\alpha}(k)$ exactly while tuning clustering and wiring length.
We use this construction to quantify which dynamical phenomena are degree-controlled (basin boundary location) and which are geometry-controlled (probability mass in long-lived non-absorbing outcomes near the boundary).

\paragraph{Testable predictions.}
The model yields structural and dynamical predictions that can be tested at several levels of stringency:
\begin{enumerate}
\item \textbf{In-degree shape as a model-selection signature (computational/connectomic):}
  The crowding rule predicts a specific one-parameter family of in-degree distributions $P_{\alpha}(k)$
  with minimum in-degree one, variance remaining bounded independently of network size,
  and a bulk location that drifts only logarithmically with network size
  (Eqs.~\eqref{eq:mean_scaling}--\eqref{eq:var_scaling}).
  In particular, the mean grows as $(1/\alpha)\log N + O_{\alpha}(1)$,
  so the peak of the histogram is expected to shift slowly to the right as $N$ increases;
  numerically, in the sparse regime, the variance saturation value is often close to $(2\alpha)^{-1}$.
  These shapes are qualitatively distinct from Erd\H{o}s--R\'enyi (Poisson), scale-free (power-law),
  or homogeneous (delta-function) degree distributions.
  Where connectomic in-degree data are available at the neuron level,
  a maximum-likelihood fit (Section~3.4) can test whether the observed histogram is consistent
  with a crowding-generated family or is better explained by alternative single-parameter models.
\item \textbf{Log-uniform wiring lengths from local encounter order (computational):}
  The prediction that spatially ordered crowding produces an approximately log-uniform
  wiring-length distribution $P(d)\propto 1/d$ and an emergent $\Prob(i\to j)\propto d^{-D}$ kernel
  (Proposition~\ref{prop:distance_kernel}) can be tested in~silico by comparing
  networks generated under competing rules (e.g.\ exponential distance decay, power-law
  preferential attachment with spatial embedding).
  Where spatially resolved connectomic data exist, the predicted $1/d$ scaling can be compared
  directly against the empirical marginal length distribution.
\item \textbf{Basin-boundary shift from degree-distribution shape (computational):}
  Networks matched in mean degree $\langle k\rangle$ but drawn from different degree families
  (crowding, ER, regular) are predicted to exhibit systematically different basin boundaries
  under threshold dynamics (Fig.~\ref{fig:baseline}).
  This is a purely in~silico prediction that distinguishes degree-shape effects from mean-degree effects.
\end{enumerate}

\paragraph{Contributions.}
(i) We provide an exact finite-$N$ recursion for $P_{\alpha}(k)$ and a generating-function recursion.
(ii) We derive rigorous scaling predictions for $\langle k\rangle$ and $\Var(k)$ as $N$ varies, including logarithmic growth of $\langle k\rangle$, bounded variance for fixed $\alpha>0$, and the asymptotic acceptance profile $p_m\sim 1/(\alpha m)$.
(iii) We show that out-degrees are approximately binomial/Poisson with mean $\langle k\rangle$ and that in- and out-degrees are weakly correlated.
(iv) We derive an HMF map $F_{\alpha}$ and a finite-size binomial-absorbing Markov approximation for basin probabilities, validating both against simulations.
(v) We compare to matched-mean Erd\H{o}s--R\'enyi and regular baselines, showing that the \emph{shape} of $P_{\alpha}(k)$ shifts basin boundaries.
(vi) We extend the crowding architecture to spatial embeddings and a shortcut-rewiring interpolation that preserve $P_{\alpha}(k)$ while generating a power-law wiring-length distribution ($P(d)\propto 1/d$) and a Kleinberg-type distance kernel ($\Prob(i\to j)\propto d^{-D}$) together with small-world structure in one and two dimensions, and we quantify the resulting dynamical deviations due to clustering.

\section{Model}

\subsection{Crowding-based wiring rule}

Consider $N$ neurons labeled $1,\dots,N$ and a directed adjacency matrix $A\in\{0,1\}^{N\times N}$ with
$A_{ij}=1$ indicating an edge $i\to j$ (no self-edges: $A_{ii}=0$).
For each target node $j$, we scan the $N-1$ candidate sources $i\neq j$ in a uniformly random order.
Let $r$ be the number of already accepted incoming edges for target $j$.
The next candidate edge is accepted with probability
\begin{equation}
\Prob(\text{accept}\mid r)=\mathrm{e}^{-\alpha r},\qquad \alpha\ge 0,
\label{eq:accept_prob}
\end{equation}
and rejected otherwise.
The exponential form is motivated by a constant decrement in log-acceptance probability per already accepted synapse: each additional afferent connection reduces $\log P$ by the same increment $\alpha$, so the unique linear choice is $\log P=-\alpha r$, i.e.\ $P=\mathrm{e}^{-\alpha r}$. No separate variational principle is assumed.
The single parameter $\alpha$ smoothly interpolates between the two extremes of network density.
Since $\mathrm{e}^{-\alpha\cdot 0}=1$, the first incoming edge is always accepted, hence every node has in-degree at least one.
In the limit $\alpha\to 0$ the graph approaches a complete directed graph (without self-edges), whereas for large $\alpha$ the typical in-degree remains $O(1)$.

\subsection{Threshold dynamics}

We define the dynamical rule here for completeness; it is first used in the heterogeneous mean-field analysis of Section~4.

Given a fixed graph $A$, we study binary states $s_i(t)\in\{-1,+1\}$ updated synchronously via
\begin{equation}
s_j(t+1)=\mathrm{sign}\!\left(\sum_{i=1}^{N} A_{ij}\,s_i(t)-\theta\right),
\label{eq:threshold_update}
\end{equation}
with threshold $\theta\in\mathbb{R}$.
Unless stated otherwise we use the tie convention $\mathrm{sign}(0)=+1$.
We study basins of attraction with respect to initial conditions with a given number of $+1$ states.

\section{Degree statistics of the crowding ensemble}

\subsection{Exact finite-$N$ recursion and generating function}

In this section, $m$ indexes the proposal step during graph construction (running from $0$ to $N{-}1$), and $r$ counts accepted incoming edges for a given target; neither should be confused with the dynamical time step $t$ of Section~2.2.

Fix a target node and let $P_m(r)$ be the probability that after processing $m$ candidate sources, the target has accepted $r$ incoming edges.
Starting from $P_0(0)=1$, the process evolves as
\begin{align}
P_{m+1}(r) &= P_m(r)\bigl(1-\mathrm{e}^{-\alpha r}\bigr)+P_m(r-1)\,\mathrm{e}^{-\alpha(r-1)},
\label{eq:Pt_recursion}
\end{align}
for $m\ge 0$ and $r\ge 0$ (with $P_m(-1)\equiv 0$).
After all candidates have been processed ($m=N-1$), the induced in-degree distribution is
$P_{\alpha}(k)=P_{N-1}(k)$.

Define the probability generating function $G_m(z)=\sum_{r\ge 0} P_m(r)\,z^r$.
From \eqref{eq:Pt_recursion} one obtains the compact recursion
\begin{equation}
G_{m+1}(z)=G_m(z)+(z-1)\,G_m(\mathrm{e}^{-\alpha}z),
\label{eq:GF_recursion}
\end{equation}
with $G_0(z)=1$.
This recursion provides a computationally efficient and conceptually transparent route to moments and scaling.

\subsection{Mean and variance scaling with system size}

Let $\mu_m=\E[r]$ and $\sigma_m^2=\Var(r)$ under $P_m$.
Exact moment recursions can be written in terms of $G_m(\mathrm{e}^{-\alpha})$ and its derivatives.
Appendix~\ref{app:moment} gives an exact inverse-process representation that yields rigorous asymptotics for $\langle k\rangle$ and $p_m$, together with a rigorous bounded-variance statement.
For fixed $\alpha>0$ and large $N$, the typical in-degree grows only logarithmically,
\begin{equation}
\langle k\rangle \sim \frac{1}{\alpha}\log N,
\label{eq:mean_scaling}
\end{equation}
while the variance remains bounded,
\begin{equation}
\Var(k)=O_\alpha(1)
\qquad (N\to\infty).
\label{eq:var_scaling}
\end{equation}
Numerically, in the sparse regime, the saturation value is often close to $(2\alpha)^{-1}$, but that constant is not obtained from an exact derivation.

\begin{figure}[t]
  \centering
  \includegraphics[width=\columnwidth]{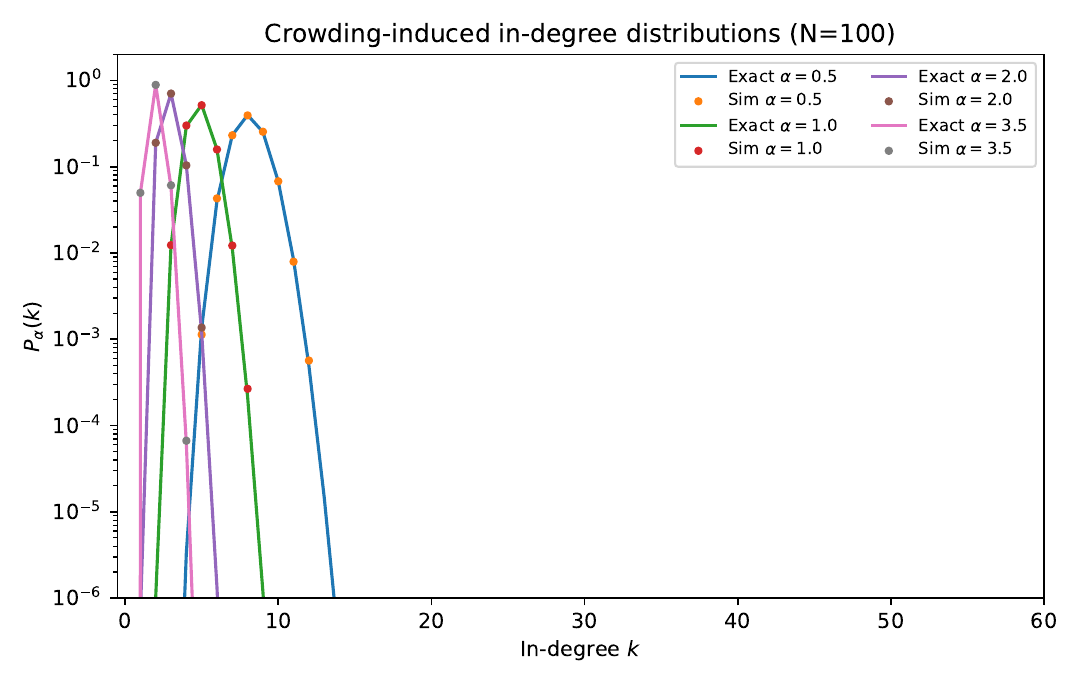}
  \caption{In-degree distributions $P_{\alpha}(k)$ for $N=100$. Solid curves show the exact finite-$N$ recursion (Eq.~\eqref{eq:Pt_recursion}); markers show simulations over independently generated graphs. Increasing $\alpha$ narrows the distribution and reduces the typical fan-in.}
  \label{fig:degdist}
\end{figure}

\begin{figure}[t]
  \centering
  \includegraphics[width=\columnwidth]{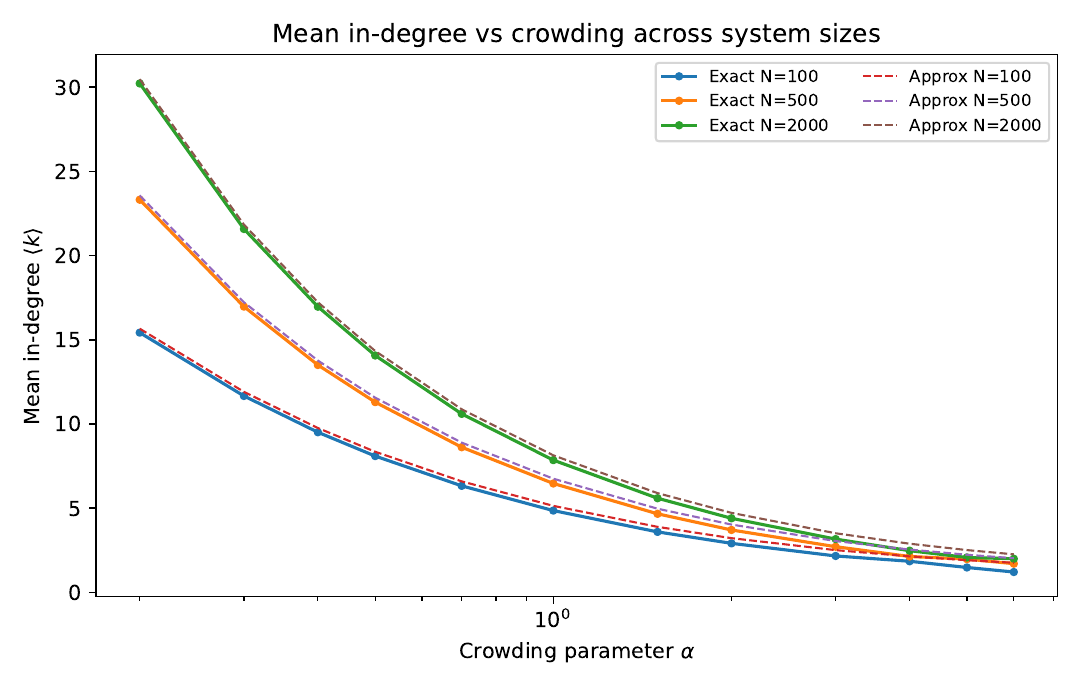}
  \caption{Mean in-degree $\langle k\rangle$ versus crowding parameter $\alpha$ for several system sizes. The crowding rule yields logarithmic growth of typical in-degree with $N$ at fixed $\alpha$ (Eq.~\eqref{eq:mean_scaling}).}
  \label{fig:mean_degree}
\end{figure}

\begin{figure}[t]
  \centering
  \includegraphics[width=\columnwidth]{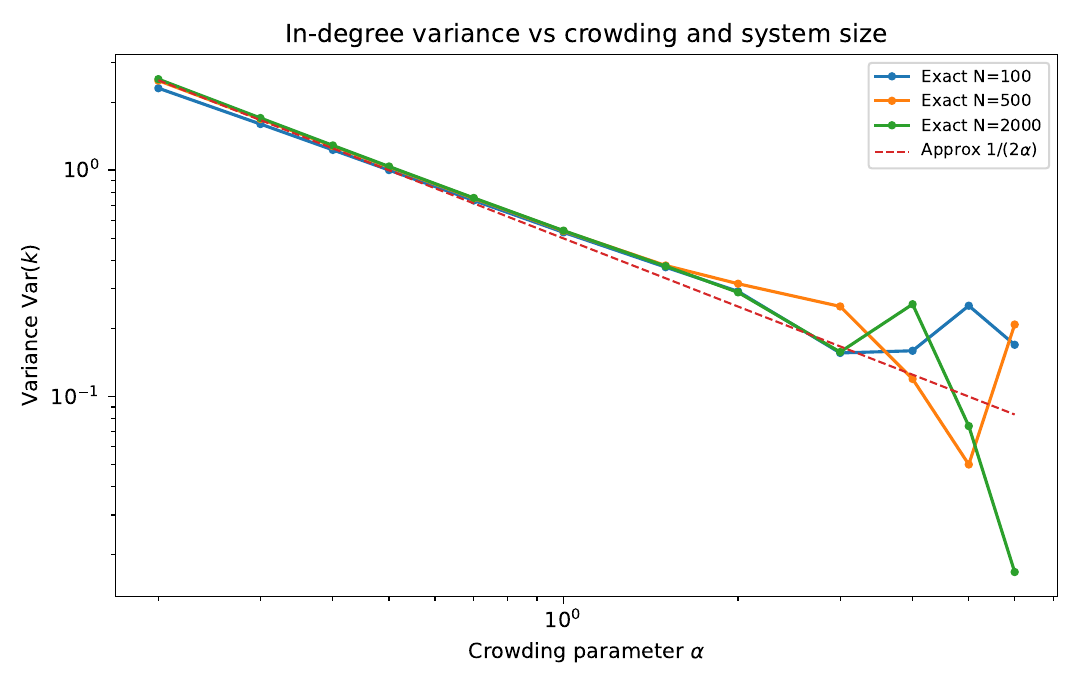}
  \caption{In-degree variance $\Var(k)$ versus crowding parameter $\alpha$ for several system sizes. The rigorous result is that $\Var(k)$ remains bounded with $N$ at fixed $\alpha$ (Eq.~\eqref{eq:var_scaling}); numerically, in the sparse regime, the saturation value is often close to $(2\alpha)^{-1}$.}
  \label{fig:var_degree}
\end{figure}

\begin{figure}[t]
  \centering
  \includegraphics[width=\columnwidth]{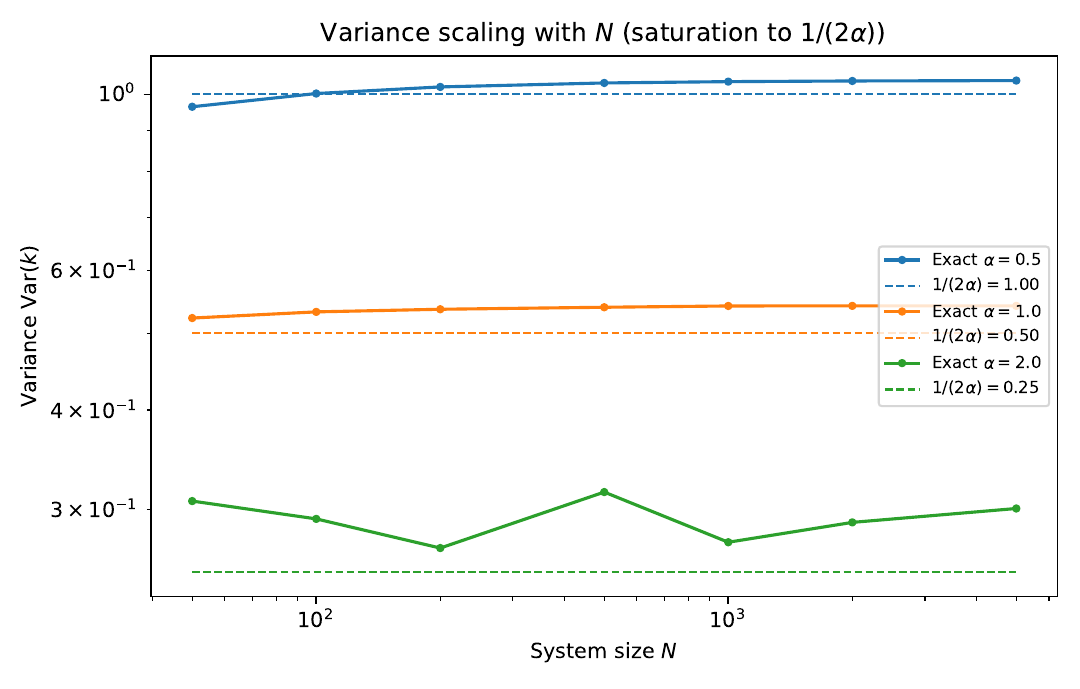}
  \caption{Variance $\Var(k)$ versus system size $N$ for representative crowding strengths, illustrating boundedness and numerical saturation with $N$ at fixed $\alpha$ in the sparse regime.}
  \label{fig:var_scaling}
\end{figure}

Figure~\ref{fig:degdist} shows representative in-degree distributions across crowding strengths.
Figure~\ref{fig:mean_degree} and figure~\ref{fig:var_degree} summarize the scaling of $\langle k\rangle$ and $\Var(k)$ with $N$ and $\alpha$, while figure~\ref{fig:var_scaling} highlights the boundedness and numerical saturation of $\Var(k)$ with $N$ at fixed $\alpha$.

\subsection{Out-degree statistics and in/out correlations}

\begin{figure}[t]
  \centering
  \includegraphics[width=\columnwidth]{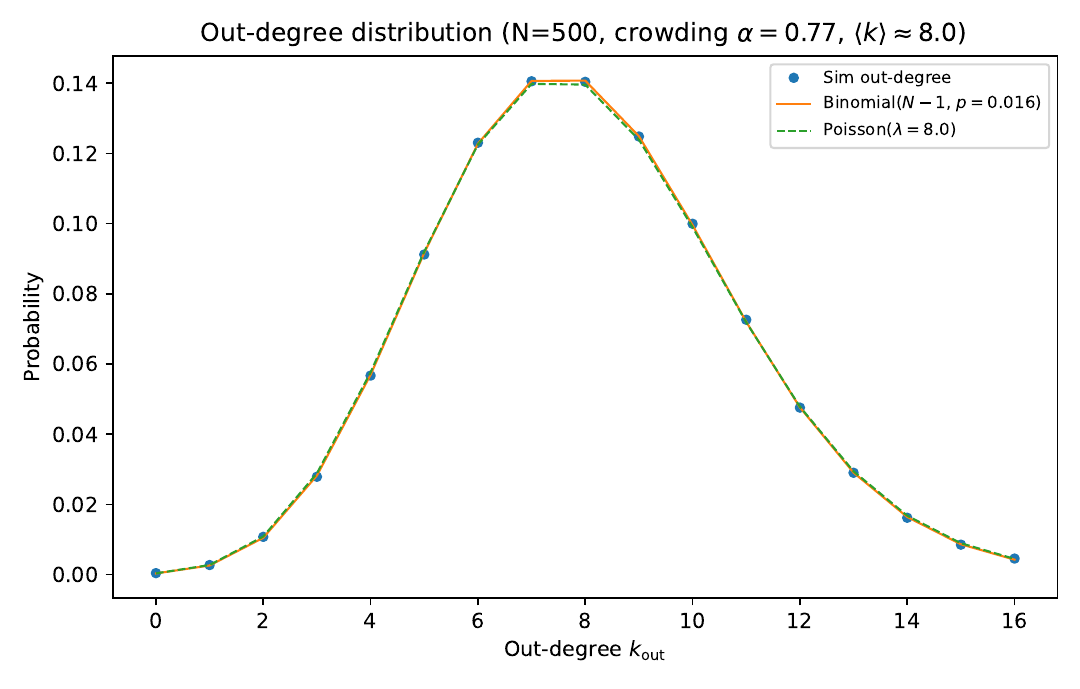}
  \caption{Out-degree distribution for the crowding ensemble at $N=500$ and $\alpha=0.77$, compared to binomial and Poisson approximations with matched mean.}
  \label{fig:outdeg}
\end{figure}

\begin{figure}[t]
  \centering
  \includegraphics[width=\columnwidth]{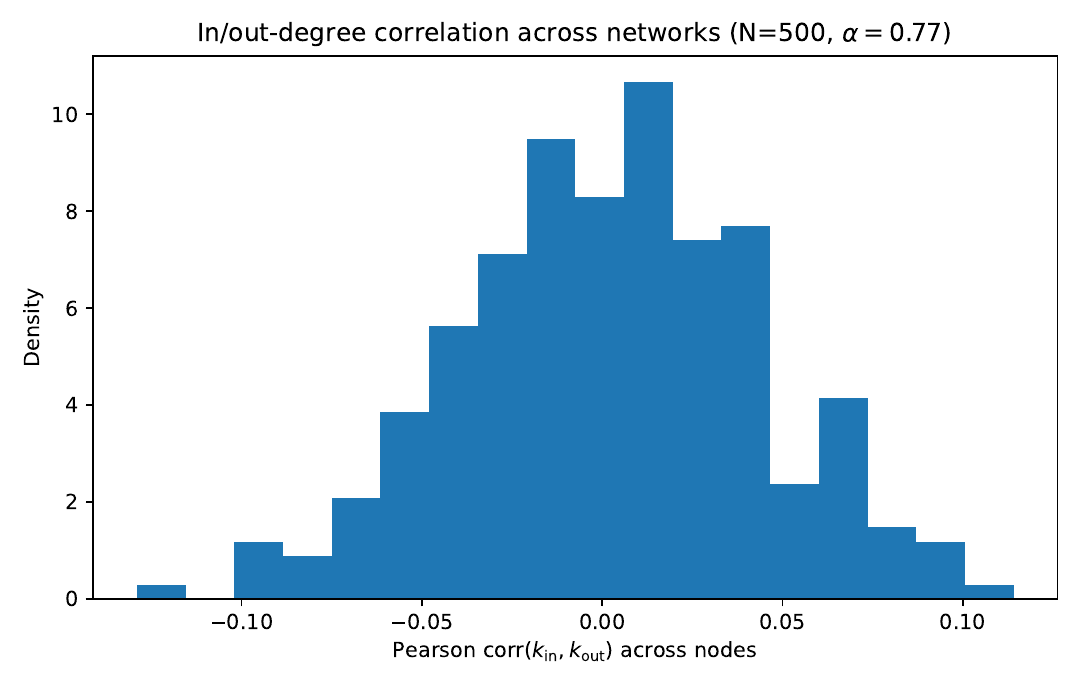}
  \caption{Histogram of Pearson correlations between node in-degree and out-degree (computed across nodes) over an ensemble of crowding graphs ($N=500$, $\alpha=0.77$).}
  \label{fig:inout_corr}
\end{figure}

In this ensemble, in-neighborhoods are generated independently across targets.
By symmetry, for any ordered pair $(i,j)$ with $i\neq j$, the edge probability is approximately
$p\approx \langle k\rangle/(N-1)$, implying that the out-degree of a node is approximately $\mathrm{Binomial}(N-1,p)$ and, in the sparse regime, approximately $\mathrm{Poisson}(\langle k\rangle)$.
Because the in-degree of a node is governed by acceptances when it is \emph{target}, while out-degree depends on being chosen as \emph{source} across many independent targets, in- and out-degrees are essentially uncorrelated (figures~\ref{fig:outdeg} and \ref{fig:inout_corr}).

\subsection{Inferring the crowding parameter from degree data}

Because $P_{\alpha}(k)$ is available exactly at finite $N$ via \eqref{eq:Pt_recursion} (or equivalently \eqref{eq:GF_recursion}), the crowding strength can be estimated from an observed in-degree histogram.
A simple approach is maximum likelihood:
given observed counts $\{n_k\}_{k\ge 1}$ with $\sum_k n_k=N$, one may estimate
$\hat{\alpha}=\arg\max_{\alpha\ge 0}\sum_{k\ge 1} n_k \log P_{\alpha}(k)$.

More explicitly, the log-likelihood of the crowding parameter is
\begin{equation}
  \mathcal{L}(\alpha)
  \;=\;
  \sum_{k\ge 1} n_k \,\log P_{\alpha}(k),
  \label{eq:loglik}
\end{equation}
where for each candidate value of $\alpha$ the probabilities
$P_{\alpha}(k)\equiv P_{N-1}(k)$ are computed exactly by iterating the
recursion~\eqref{eq:Pt_recursion} from $m=0$ to $m=N-1$.
Because the recursion involves $O(N)$ terms at each step, the cost of
a single likelihood evaluation is $O(N^2)$, which is negligible for
the system sizes considered here.
In practice $\mathcal{L}(\alpha)$ is a smooth,
unimodal function of $\alpha$ on $(0,\infty)$, so the
maximum-likelihood estimate
$\hat{\alpha}=\arg\max_{\alpha}\mathcal{L}(\alpha)$ can be located
efficiently by any standard scalar optimizer (e.g.\ Brent's method or
golden-section search over a coarse bracket obtained from an initial
grid evaluation).
An approximate confidence interval for $\hat{\alpha}$ is available
from the observed Fisher information:
$\hat{\alpha}\pm z_{1-\gamma/2}/\!\sqrt{I(\hat{\alpha})}$,
where $I(\hat{\alpha})=-\mathcal{L}''(\hat{\alpha})$ is estimated by
finite differences.

\begin{figure}[t]
  \centering
  \includegraphics[width=\columnwidth]{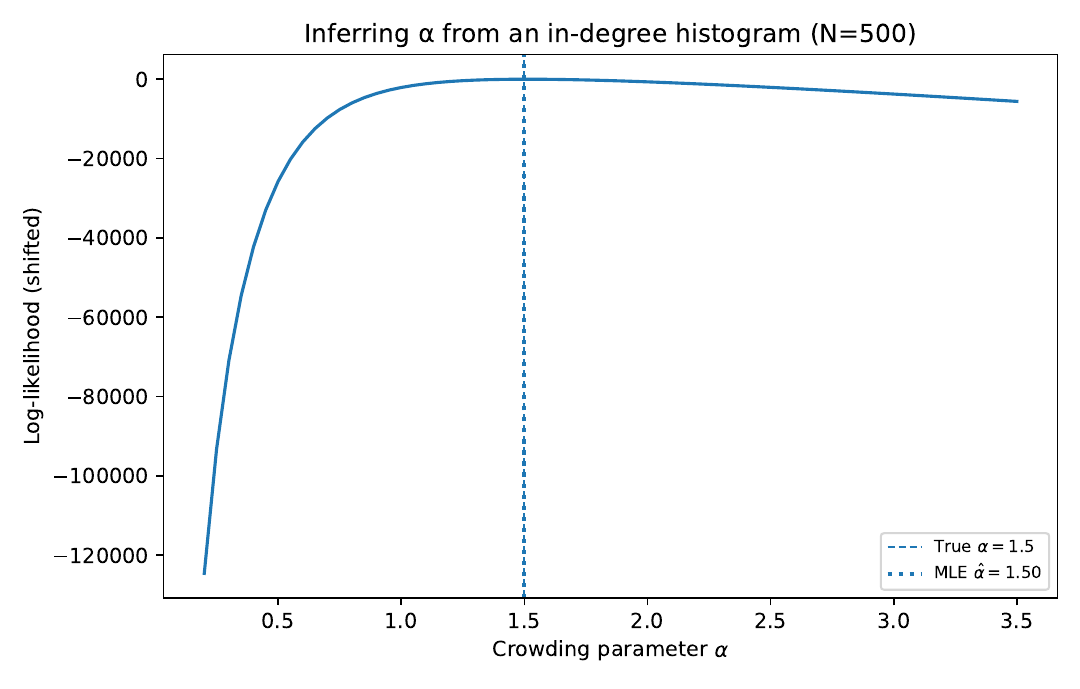}
  \caption{Maximum-likelihood inference of the crowding parameter $\alpha$ from an observed in-degree histogram. Synthetic experiments show accurate recovery in regimes where $P_{\alpha}(k)$ is informative.}
  \label{fig:alpha_infer}
\end{figure}

Figure~\ref{fig:alpha_infer} illustrates that $\alpha$ can be recovered accurately from synthetic data in regimes where the histogram is sufficiently informative (moderate $N$ and $\alpha$ away from degenerate limits).

\section{Heterogeneous mean-field dynamics}

Following mean-field / statistical-neurodynamic ideas \cite{Amari1971,Amari1972,Amari1974,Amari1977}, we assume that (i) node states are i.i.d.\ with $\Prob(s=+1)=x(t)$ and (ii) edges are annealed except for their in-degree $k$.
Under these assumptions, the probability that a node with in-degree $k$ is active at the next step is
\begin{equation}
F_k(x;\theta)=\sum_{\ell=\lceil (k+\theta)/2\rceil}^{k} \binom{k}{\ell}\,x^\ell(1-x)^{k-\ell}.
\label{eq:Fk}
\end{equation}
Averaging over the crowding-induced in-degree distribution gives the ensemble map
\begin{equation}
x(t+1)=F_{\alpha}\bigl(x(t);\theta\bigr):=\sum_{k\ge 1} P_{\alpha}(k)\,F_k(x(t);\theta).
\label{eq:Falpha}
\end{equation}

\begin{figure}[t]
  \centering
  \includegraphics[width=\columnwidth]{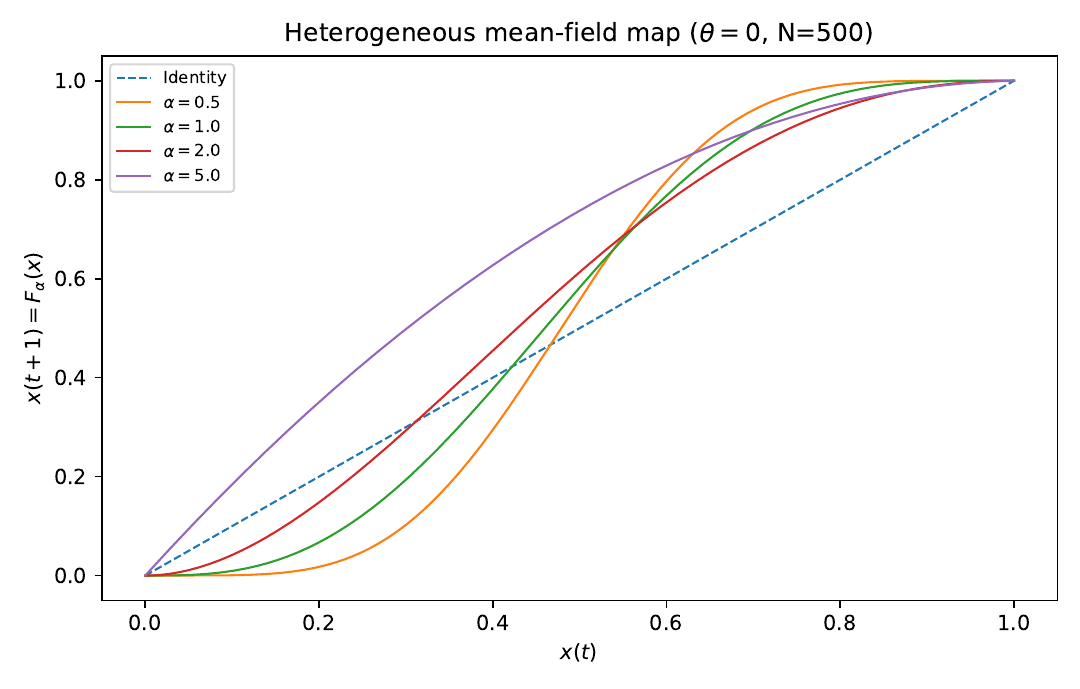}
  \caption{Heterogeneous mean-field map $x(t+1)=F_{\alpha}(x(t);\theta)$ (Eq.~\eqref{eq:Falpha}) for representative crowding strengths at $\theta=0$. The induced degree distribution $P_{\alpha}(k)$ deforms the macroscopic map and shifts unstable fixed points that govern basin boundaries.}
  \label{fig:hmf_map}
\end{figure}

Figure~\ref{fig:hmf_map} illustrates how changing $\alpha$ (and therefore the full distribution $P_{\alpha}(k)$) deforms the macroscopic map $F_{\alpha}$.

For intermediate crowding strengths where $P_{\alpha}(k)$ concentrates near small $k$ (e.g.\ $k\approx 2$), the map can lie above the identity line on most of $(0,1]$, implying dominance of the all-$+1$ state from nearly any nonzero initial activity.
For weaker crowding (smaller $\alpha$, larger typical $k$), an interior unstable fixed point emerges near $x\approx 1/2$, predicting a sharper basin boundary as $N\to\infty$.
For very strong crowding (large $\alpha$, $k\approx 1$), $F_{\alpha}$ approaches the identity map, foreshadowing stronger finite-size effects and an increased prevalence of short cycles.

\section{Finite-size basin probabilities via an absorbing Markov closure}

The deterministic map \eqref{eq:Falpha} predicts fixed points and, in large $N$, basin boundaries via unstable fixed points.
To estimate finite-size basin probabilities we follow a standard ``annealed'' closure:
given $x(t)=a/N$ where $a$ is the number of currently active nodes, each node updates independently with probability $F_{\alpha}(a/N;\theta)$, so the active count evolves as
\begin{equation}
R(t+1)\,\big|\,R(t)=a\ \sim\ \mathrm{Binomial}\!\left(N,\ F_{\alpha}(a/N;\theta)\right).
\label{eq:binom_chain}
\end{equation}
For parameter regimes with $F_{\alpha}(0;\theta)=0$ and $F_{\alpha}(1;\theta)=1$---in particular, for the cases emphasized below, $\theta=0$ and also $\theta=1$ under the convention $\mathrm{sign}(0)=+1$---$R=0$ and $R=N$ act as absorbing macrostates.
Let $u_a=\Prob(\text{hit }N\text{ before }0\mid R(0)=a)$ be the committor/hitting probability.
Then $u_0=0$, $u_N=1$, and for $1\le a\le N-1$,
\begin{equation}
\begin{aligned}
u_a &= \sum_{n=0}^{N} \binom{N}{n}\, q_a^n(1-q_a)^{N-n}\,u_n,\\
&\qquad q_a:=F_{\alpha}(a/N;\theta).
\end{aligned}
\label{eq:committor}
\end{equation}
For moderate $N$, \eqref{eq:committor} is solved as a linear system; for larger $N$ we estimate $u_a$ accurately by Monte Carlo simulation of the chain \eqref{eq:binom_chain}.
This construction is closely related to committor functions for stochastic dynamical systems \cite{LindnerHellmann2019}.

\begin{figure}[t]
  \centering
  \includegraphics[width=\columnwidth]{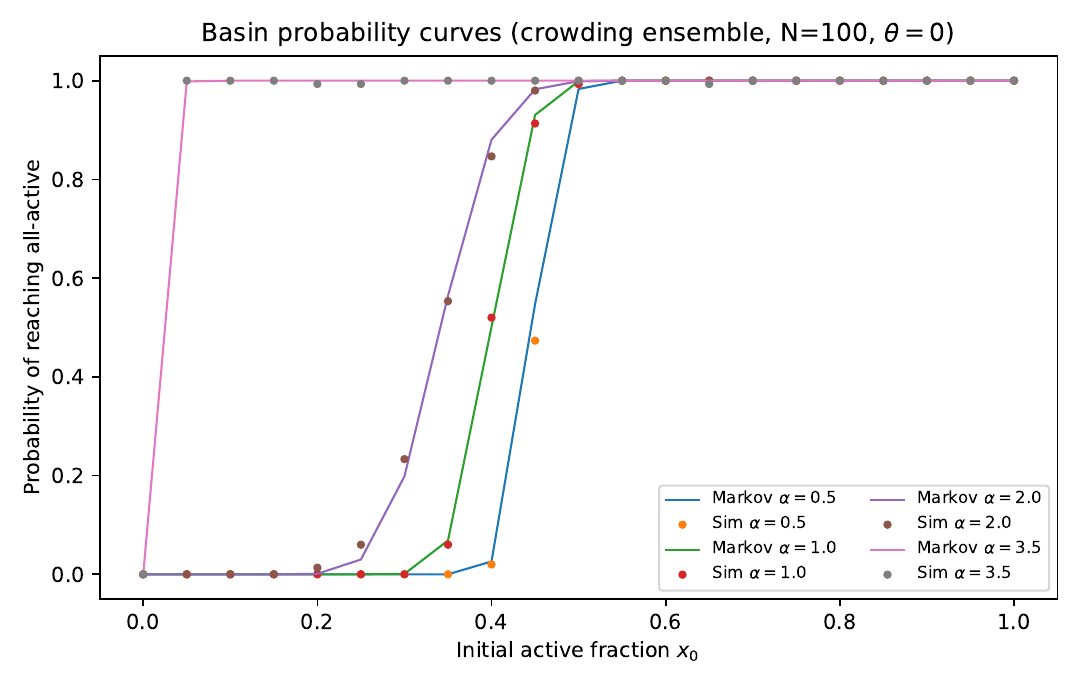}
  \caption{Basin (hitting) probability of reaching the all-$+1$ state versus initial active fraction for $N=100$ and $\theta=0$. Solid curves show the absorbing-Markov committor prediction (Eq.~\eqref{eq:committor}); markers show direct simulations of the threshold dynamics on crowding graphs.}
  \label{fig:basin_N100}
\end{figure}

\begin{figure}[t]
  \centering
  \includegraphics[width=\columnwidth]{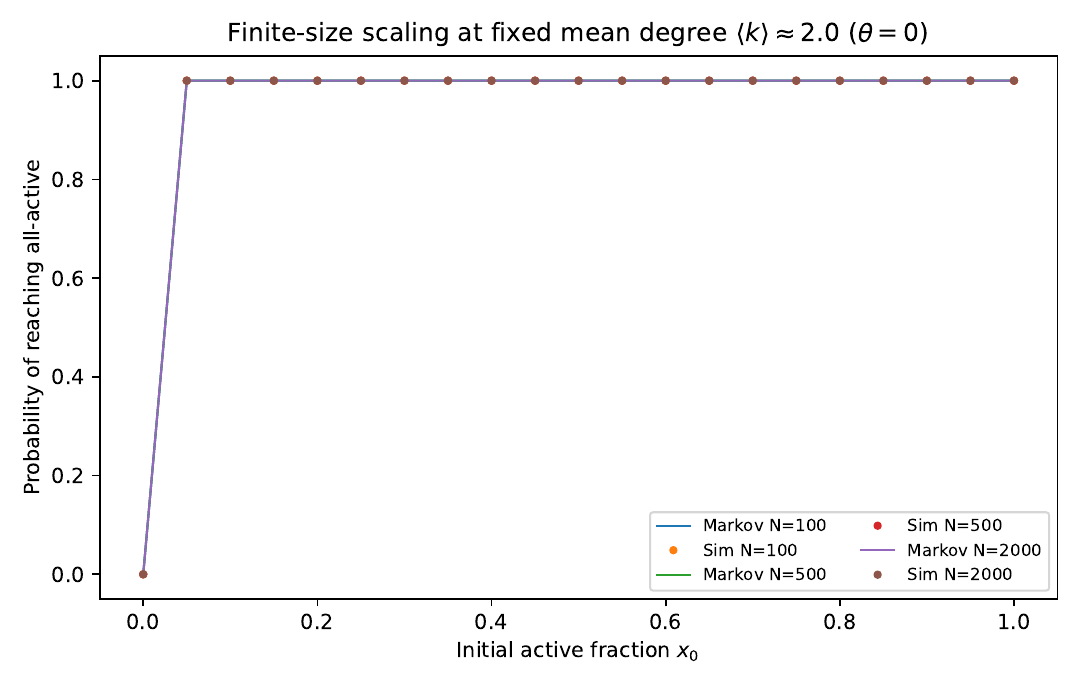}
  \caption{Finite-size scaling of basin curves at matched mean degree across system sizes. The absorbing-Markov closure captures the sharpening of the basin boundary with increasing $N$ while revealing systematic finite-size deviations near the critical initial activity.}
  \label{fig:finite_scaling}
\end{figure}

\begin{figure}[t]
  \centering
  \includegraphics[width=\columnwidth]{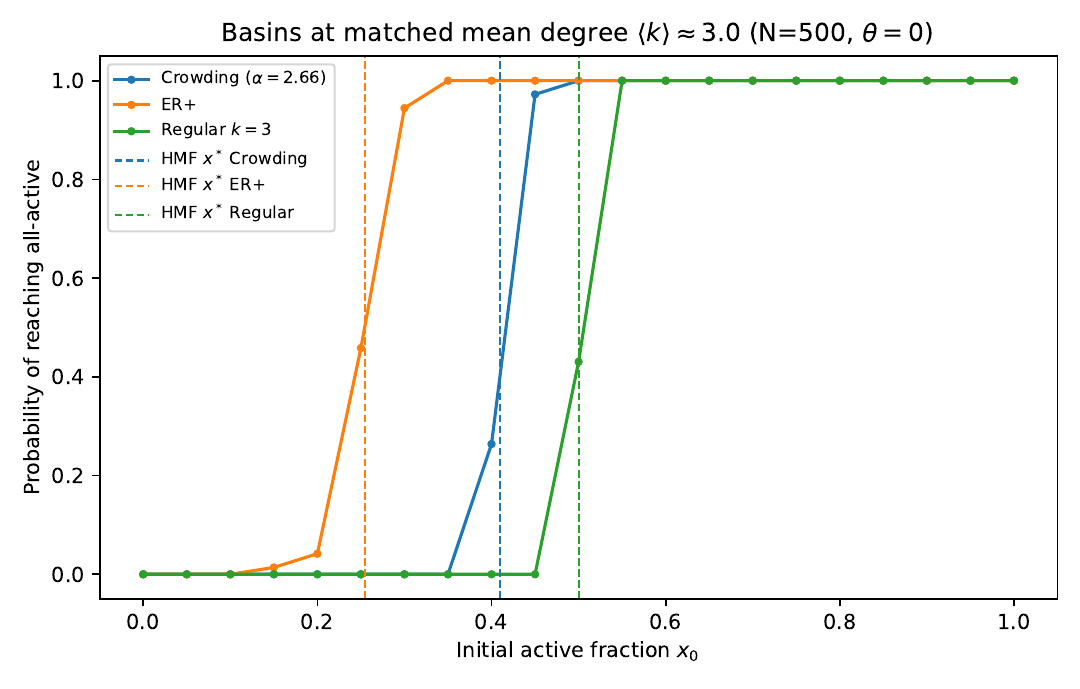}
  \caption{Baseline comparison at matched mean degree $\langle k\rangle\approx 3$ ($N=500$, $\theta=0$): basin curves for the crowding ensemble, an ER$^{+}$ graph (a directed Erd\H{o}s--R\'enyi baseline conditioned to have minimum in-degree 1), and a regular in-degree graph. Vertical dashed lines indicate HMF unstable fixed points for each degree distribution.}
  \label{fig:baseline}
\end{figure}

\begin{figure}[t]
  \centering
  \includegraphics[width=\columnwidth]{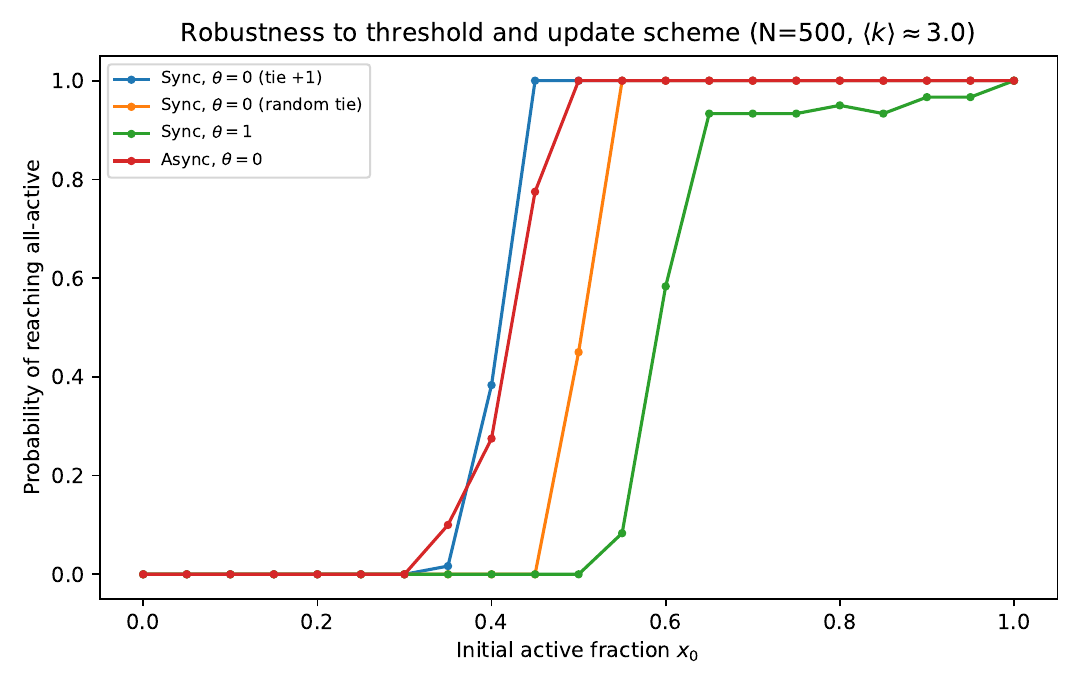}
  \caption{Robustness of simulated basin curves in the crowding ensemble ($N=500$, $\langle k\rangle\approx 3$) to tie-breaking at $\theta=0$, changing the threshold to $\theta=1$, and asynchronous updates.}
  \label{fig:robustness}
\end{figure}

Figure~\ref{fig:basin_N100} compares \eqref{eq:committor} to direct simulations of \eqref{eq:threshold_update} on graphs from the crowding ensemble, showing accurate basin-probability estimates over a broad range of $\alpha$.
We also report finite-size scaling at fixed mean degree, comparisons to matched-mean Erd\H{o}s--R\'enyi and regular baselines, and robustness to threshold choice, asynchronous updates, and tie-breaking (figures~\ref{fig:finite_scaling}--\ref{fig:baseline} and \ref{fig:robustness}). In Fig.~\ref{fig:baseline}, ER$^{+}$ denotes a directed Erd\H{o}s--R\'enyi baseline conditioned to have minimum in-degree one, so that it shares the support constraint $k\ge 1$ with the crowding ensemble.

\section{Spatially ordered crowding and small-world structure}

The wiring rule in Eq.~\eqref{eq:accept_prob} is agnostic to the \emph{order} in which candidate sources are proposed.
This makes it straightforward to incorporate spatial embedding without changing the analytically derived degree statistics.

\begin{figure}[t]
  \centering
  \includegraphics[width=\columnwidth]{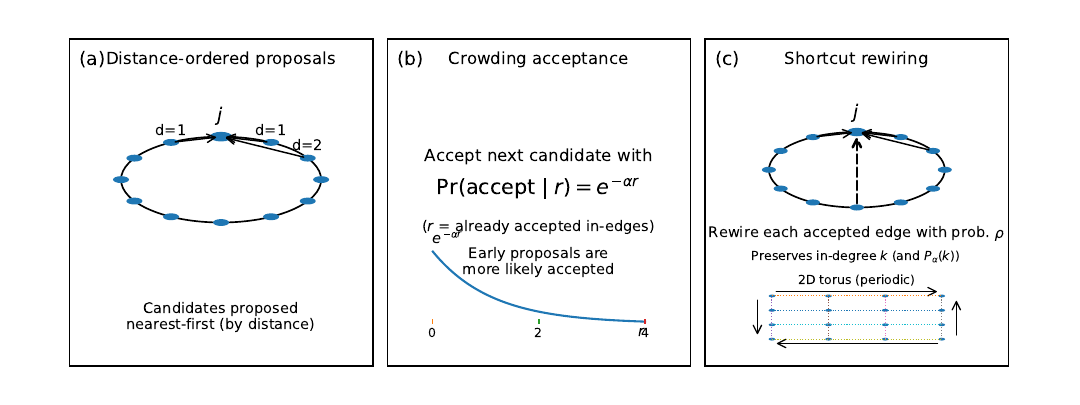}
  \caption{Schematic construction of spatially ordered crowding and the controlled small-world interpolation.
  (a) Nodes are embedded in space (ring shown) and candidates for each target $j$ are proposed nearest-first, with random tie-breaking inside equal-distance shells.
  (b) The crowding acceptance rule depends only on the current accepted in-degree $r$: later synapses are less likely to be accepted.
  (c) A small-world interpolation is obtained by rewiring each accepted edge independently with probability $\rho$ to a random source not already connected to the same target (excluding self-edges), preserving each target's in-degree and hence $P_{\alpha}(k)$.
  The same distance-ordering idea extends to periodic $D$-dimensional lattices (2D torus schematic).}
  \label{fig:smallworld_schematic}
\end{figure}

\subsection{Distance-ordered proposals preserve $P_{\alpha}(k)$}

Place nodes on a ring and, for each target $j$, propose candidates in increasing ring distance (nearest neighbours first).
Candidates at the same distance are randomly permuted within each equal-distance shell.
Run the same acceptance rule \eqref{eq:accept_prob}.
Because acceptance depends only on the current accepted count $r$, the distribution of the \emph{number} of accepted edges is invariant under any permutation of the candidate list.
Therefore, the in-degree distribution remains exactly $P_{\alpha}(k)$ even though the \emph{geometry} of selected edges changes.

\begin{figure}[t]
  \centering
  \includegraphics[width=\columnwidth]{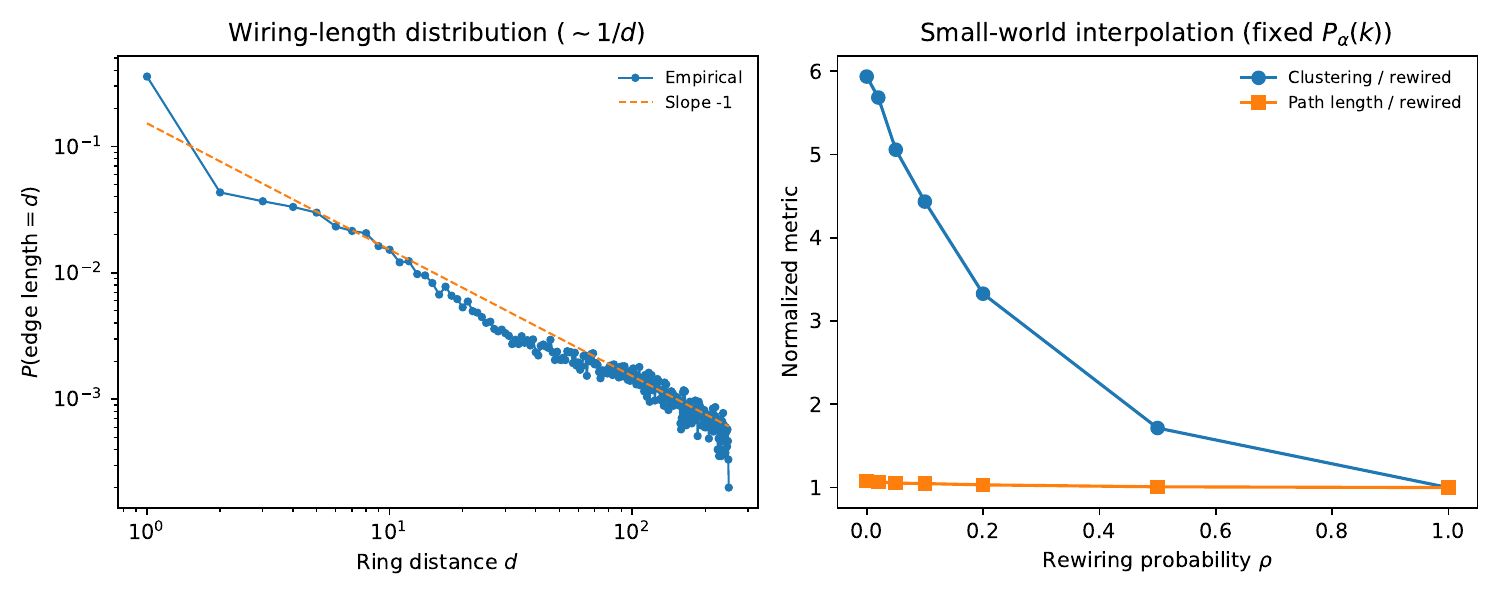}
  \caption{Spatially ordered crowding yields a power-law wiring-length distribution and small-world structure at fixed degree statistics ($N=500$, $\alpha=2.66$).
  (a) Edge-length distribution on a ring is well approximated by $P(d)\propto d^{-1}$.
  (b) Interpolating from spatially ordered graphs ($\rho=0$) to fully rewired graphs ($\rho=1$) preserves $P_{\alpha}(k)$ while reducing clustering.
  Shown are the mean local clustering coefficient and mean shortest-path length of the symmetrized simple graph (averaged over connected node pairs), normalized by the fully rewired ($\rho=1$) baseline.}
  \label{fig:smallworld_structure}
\end{figure}

A useful quantity is the acceptance probability of the $m$-th proposal in the ordered list:
\begin{equation}
p_m=\E\!\left[\mathrm{e}^{-\alpha R_{m-1}}\right]=G_{m-1}(\mathrm{e}^{-\alpha}),
\label{eq:pt_accept_rank}
\end{equation}
where $R_{m-1}$ is the accepted count after $m-1$ proposals and $G_m$ is the generating function in Eq.~\eqref{eq:GF_recursion}.
Appendix~\ref{app:moment} proves the large-$m$ asymptotic
\begin{equation}
p_m \sim \frac{1}{\alpha\,m}.
\label{eq:pt_asymptotic}
\end{equation}
In a distance-ordered list on a ring, the proposal rank $m$ is proportional to distance $d$.
Equation~\eqref{eq:pt_asymptotic} therefore predicts an approximately power-law wiring-length distribution $P(\text{edge length}=d)\propto 1/d$, i.e.\ a roughly constant expected number of edges per logarithmic distance scale.
Figure~\ref{fig:smallworld_structure}a confirms this $1/d$ scaling numerically.
This heavy-tailed length profile is robust to ``soft'' distance ordering in which proposals are mostly local but occasionally drawn uniformly at random (Appendix~\ref{app:softbias}, Fig.~\ref{fig:softbias}).

\begin{proposition}[Emergent distance kernel from crowding and spatial ordering]
\label{prop:distance_kernel}
Consider a spatial embedding in $D$ dimensions and a candidate list ordered by increasing distance from a target node.
Under the crowding rule in Eq.~\eqref{eq:accept_prob}, the acceptance probability of proposal rank $m$ satisfies $p_m\sim (\alpha m)^{-1}$ for large $m$.
If the number of candidates within distance $d$ scales as $m\propto d^{D}$, then the induced pairwise connection probability obeys $\Prob(i\to j)\propto d^{-D}$.
Consequently, the marginal wiring-length distribution scales as $P(d)\propto 1/d$, i.e., the expected number of accepted edges per logarithmic distance bin is approximately constant.
\end{proposition}

The same argument extends to higher-dimensional spatial embeddings.
On a $D$-dimensional periodic lattice (torus), the number of candidate sources within distance $d$ grows as $d^{D}$, so the proposal rank corresponding to distance $d$ satisfies $m\propto d^{D}$.
Combining this with Eq.~\eqref{eq:pt_asymptotic} implies an approximate pairwise connection probability $\Prob(i\to j)\propto d^{-D}$, matching the exponent that plays a distinguished role in Kleinberg-type navigable small-world models.
Multiplying by the surface-area growth ($\propto d^{D-1}$) yields an approximately dimension-independent marginal wiring-length distribution $P(d)\propto 1/d$.
Appendix~\ref{app:torus2d} confirms these predictions numerically on a two-dimensional torus.

\subsection{Shortcut rewiring and a controlled small-world interpolation}

To obtain a canonical small-world interpolation, we post-process a distance-ordered crowding graph by rewiring each accepted edge independently with probability $\rho$ to a uniformly random source that is not already connected to the same target (excluding self-edges, with resampling if necessary).
This changes \emph{which} sources connect to a target but not \emph{how many} connect, so $P_{\alpha}(k)$ is preserved by construction.
For the structural diagnostics in Figs.~\ref{fig:smallworld_structure}b and \ref{fig:torus2d_structure}b, we work with the symmetrized simple graph obtained by replacing $A$ with $A\lor A^{\mathsf T}$; the clustering coefficient is the mean local clustering coefficient of this symmetrized graph, and the characteristic path length is the mean shortest-path distance over connected node pairs. The plotted values are normalized by the corresponding $\rho=1$ values.
As $\rho$ increases, clustering decreases toward the random baseline while path length remains short (Fig.~\ref{fig:smallworld_structure}b), yielding a ``high-clustering/short-path'' regime characteristic of small-world networks \cite{WattsStrogatz1998,SpornsZwi2004,BassettBullmore2006,Kleinberg2000}.

\subsection{Dynamical consequences: basin curves versus non-absorbing outcomes}

\begin{figure}[t]
  \centering
  \includegraphics[width=\columnwidth]{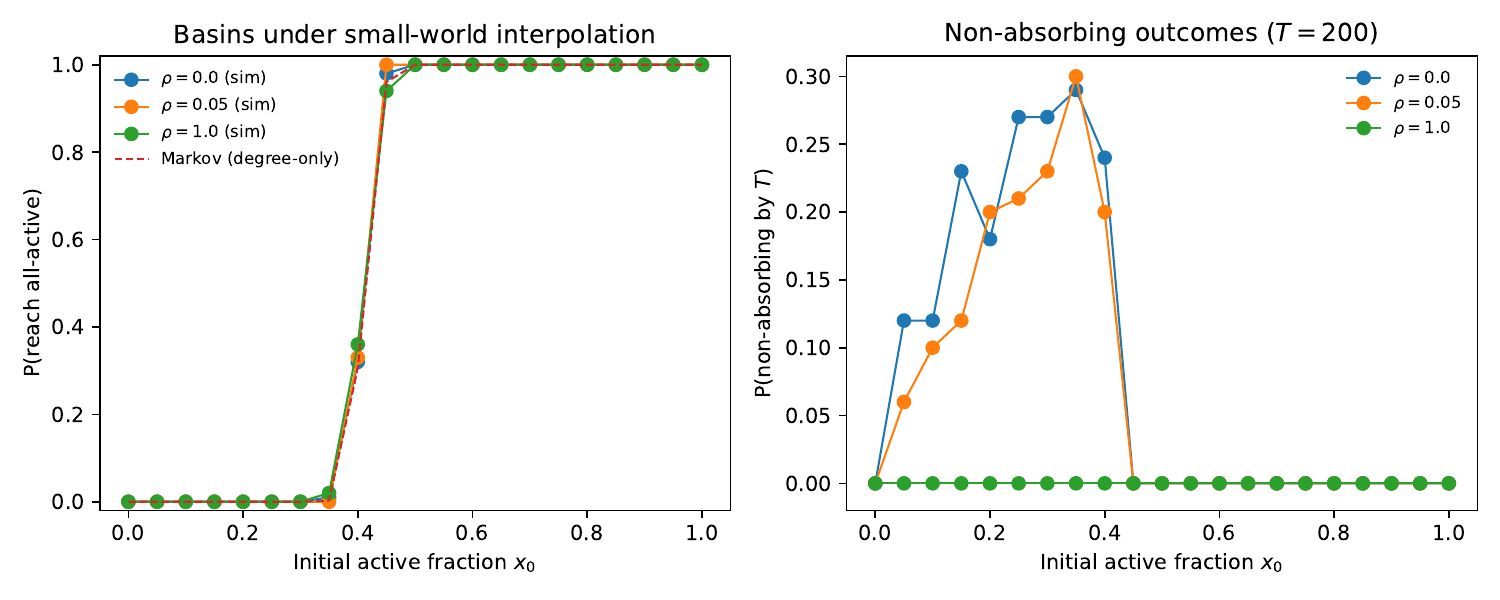}
  \caption{Small-world effects on threshold dynamics at fixed degree statistics ($N=500$, $\alpha=2.66$, $\theta=0$).
  (a) Basin curves $\Pr\{\text{reach all-active}\}$ as a function of initial activity $x_0$ for spatially ordered graphs ($\rho=0$), a weakly rewired small-world ($\rho=0.05$), and fully rewired graphs ($\rho=1$), compared to the degree-only absorbing-Markov prediction.
  (b) Probability that dynamics does not reach either absorbing state within $T=200$ synchronous updates, showing that local clustering increases the prevalence of long-lived non-absorbing outcomes near the basin boundary.}
  \label{fig:smallworld_dynamics}
\end{figure}

The absorbing-Markov closure (Section~4) depends only on $P_{\alpha}(k)$ and therefore predicts the \emph{same} basin curve for all $\rho$ at fixed $(N,\alpha,\theta)$.
Simulations confirm that the probability of reaching the all-active state is largely stable across the small-world interpolation (Fig.~\ref{fig:smallworld_dynamics}a), supporting the central message that degree statistics strongly constrain the basin boundary.

However, spatially ordered graphs exhibit a substantial probability of \emph{non-absorbing} outcomes under synchronous updates near the basin boundary (Fig.~\ref{fig:smallworld_dynamics}b), consistent with long-lived transients, short cycles, or mixed fixed points enabled by local motifs.
These outcomes are suppressed by shortcut rewiring and are essentially absent in the fully rewired case.
Consequently, the two-absorbing-state Markov closure is most accurate when the network is sufficiently ``well-mixed'' (large $\rho$), while deviations at $\rho\approx 0$ can be attributed to probability mass leaking into long-lived non-absorbing outcomes---consistent with candidate additional attractors---that are not represented in the closure.

\subsection{Biological interpretation}

Empirical connectomes and functional networks often display small-world signatures: high clustering coexisting with short characteristic path length \cite{SpornsZwi2004,BassettBullmore2006}.
Within the present framework, $\alpha$ primarily controls fan-in statistics, while spatial ordering and shortcut density control wiring length and clustering.
The predicted $1/d$ wiring-length distribution provides a minimal mechanism by which a local growth rule can generate multi-scale connectivity, potentially relevant to how predominantly local synaptic growth coexists with a minority of long-range projections in cortex.

\paragraph{Crowding, spatial embedding, and small-world structure.}
We stress that the small-world property---high clustering coexisting with short characteristic path length---is not a direct consequence of the crowding rule alone.
Small-worldness generically requires two ingredients: (i) predominantly local wiring that produces high clustering, and (ii) a mechanism for long-range shortcuts that keeps path lengths short \cite{WattsStrogatz1998}.
In the present model, spatial ordering of candidate encounters supplies ingredient~(i), while the broad tail of the crowding-induced acceptance profile ($p_m\sim 1/(\alpha m)$; Eq.~\eqref{eq:pt_asymptotic}) naturally generates a minority of long-range edges that serve as shortcuts, supplying ingredient~(ii).
The specific contribution of the crowding mechanism is therefore not small-worldness per se, but rather the \emph{emergent distance kernel}: the $1/d$ (log-uniform) wiring-length distribution and the Kleinberg-type pairwise connection probability $\Prob(i\to j)\propto d^{-D}$ arise without imposing any distance-dependent connection law.
This is distinct from models that achieve small-worldness by prescribing a rewiring probability (Watts--Strogatz) or an explicit distance kernel (Kleinberg), and provides a developmental mechanism---crowding-limited synaptogenesis combined with proximity-biased encounter order---that produces these statistical regularities as a byproduct of local resource constraints.

\section{Discussion}

\subsection{Summary: an analytically tractable crowding ensemble}

We introduced a minimal degree-dependent wiring rule motivated by synaptic crowding and derived exact finite-$N$ in-degree statistics, compact generating-function recursions, and simple scaling laws for mean and variance.
A notable feature is that the crowding parameter $\alpha$ provides a transparent ``fan-in knob'': for fixed $\alpha>0$ the typical in-degree grows only logarithmically with system size, while the variance remains bounded (and numerically saturates) in the sparse regime.
Out-degree statistics are close to binomial/Poisson with mean $\langle k\rangle$ and show little correlation with in-degree, reflecting the independence of in-neighborhood construction across targets.

\subsection{Degree statistics control basin boundaries in a threshold network}

To connect structure to dynamics, we analyzed synchronous threshold dynamics on this ensemble.
A heterogeneous mean-field (HMF) map, averaged over the \emph{induced} in-degree distribution $P_{\alpha}(k)$, predicts the location of unstable fixed points that act as macroscopic basin boundaries.
To capture finite-size effects we introduced a one-dimensional ``binomial absorbing Markov'' closure for the active-count process and computed committor (hitting) probabilities.
Across a broad range of $(N,\alpha)$, these degree-based predictions agree well with direct simulations.

Matched-baseline comparisons highlight that it is not enough to match $\langle k\rangle$ alone: the \emph{shape} of $P_{\alpha}(k)$ changes the HMF map and shifts the basin boundary.
This provides a concrete example in which a mechanistic wiring rule induces a specific degree distribution that leaves a measurable dynamical fingerprint.

\subsection{Spatial embedding separates the roles of degrees and clustering}

The crowding rule depends only on the current accepted in-degree, so the distribution of the \emph{number} of accepted edges is invariant under permutations of the candidate list.
This invariance enables a spatial embedding in which candidates are proposed in order of distance, without altering $P_{\alpha}(k)$.
Notably, this construction does not impose any distance-dependent connection probability; the distance dependence arises from the interaction between a rank-dependent acceptance probability and the geometry of the candidate ordering.
A simple rank-based argument predicts that, for large proposal rank $m$, the acceptance probability scales as $p_m\sim 1/(\alpha m)$.
In spatial orderings where proposal rank grows with the volume of a ball, this yields a Kleinberg-type distance-dependent pairwise connection probability \cite{Kleinberg2000} and an approximately power-law wiring-length distribution $P(d)\propto 1/d$ (Fig.~\ref{fig:smallworld_structure}a and Appendix~\ref{app:torus2d}).

We then introduced a shortcut-rewiring interpolation that preserves $P_{\alpha}(k)$ by construction while tuning clustering and path length.
In this setting, basin curves for the all-inactive/all-active absorbing outcomes remain largely stable across the interpolation, consistent with the degree-only nature of the HMF and absorbing-Markov closures.
In contrast, strongly spatial (high-clustering) networks exhibit a substantial probability of \emph{non-absorbing} outcomes near the basin boundary under synchronous updates (Fig.~\ref{fig:smallworld_dynamics}b).
These outcomes are suppressed by rewiring and are essentially absent in the fully rewired case, indicating that local motifs and clustering primarily modulate the prevalence of long-lived non-absorbing outcomes near the basin boundary.
Practically, this clarifies when degree-only (annealed) basin estimates can be expected to work well (sufficient mixing) and when they can fail (strong local structure).

\subsection{Implications for biological networks and learning}

Empirical connectomes and functional networks often display small-world signatures \cite{SpornsZwi2004,BassettBullmore2006}.
Within the present framework, $\alpha$ controls fan-in statistics while spatial ordering and shortcut density control wiring length and clustering.
The emergent $1/d$ wiring-length distribution implies a roughly constant expected number of synapses per logarithmic distance scale, providing a minimal mechanism by which predominantly local growth can coexist with a minority of long-range projections.
Although our spatial model is stylized, it offers a tractable bridge between degree regulation (homeostatic constraints) and multi-scale network organization.

From the perspective of modern machine learning, the crowding mechanism can be viewed as a \emph{structural prior} that regulates per-unit fan-in.
Dynamic sparse training methods---including Deep Rewiring \cite{Bellec2018DeepRewiring}, Sparse Evolutionary Training \cite{Mocanu2018SET}, and RigL \cite{Evci2020RigL}---suggest that changing connectivity under a fixed sparsity budget can improve optimization and generalization.
The present ensemble provides an analytically controllable family of sparse directed backbones on which one can study how fan-in regulation and motif structure interact with learning rules, potentially complementing weight-centric analyses.

\subsection{Limitations and future directions}

We focused on a simplified setting: unweighted excitatory edges, homogeneous thresholds, and an annealed approximation for basin probabilities that collapses network dynamics to a one-dimensional active-count process.
Spatial effects were introduced through deterministic distance ordering and post-hoc shortcut rewiring rather than through an explicit growth-and-cost optimization.
Future extensions include (i) signed weights and inhibition, (ii) heterogeneous thresholds and external fields, (iii) coupling structural rewiring to activity and synaptic plasticity, and (iv) richer spatial models (continuous space, anisotropy, explicit wiring costs) that retain the analytical separability between degree statistics and geometry.

\section*{Methods}

\subsection*{Numerical experiments and reproducibility}

All numerical results reported in this manuscript (Monte Carlo graph generation, threshold-dynamics simulations, and small-world / spatial-embedding analyses) are reproducible from the accompanying Jupyter notebook \textit{generate\_figures.ipynb} provided with this submission as \textbf{Supplementary Software 1}.
Running the notebook generates the 17 figure PDF files that are included in the manuscript source (Figs.~1--17) and reproduces the simulation curves shown in the main text.
A fixed pseudorandom seed is used by default (see \texttt{rng = default\_rng(1)} in Supplementary Software 1), and the notebook collects all Monte Carlo sample-size parameters in a single \texttt{SIM\_SETTINGS} dictionary for transparency.

\subsection*{Large language model usage}

LLM tools (including GPT-based systems) were used to assist with drafting, language editing, code organization, and exploratory mathematical derivations. All equations, analyses, and scientific conclusions were independently verified and are the responsibility of the author.

\begin{acknowledgments}
A portion of this work was originally submitted as the author's undergraduate thesis in the Department of Physics at Waseda University.
The author gratefully acknowledges Professor Yoji Aizawa for supervision during that period.
Subsequent analyses and extensions were carried out independently by the author. The author's current affiliation is provided for identification purposes only.
\end{acknowledgments}

\section*{Data availability}
This study is theoretical and computational. No external datasets were analyzed.
All data points underlying the figures can be regenerated by running the accompanying figure-generation notebook (Supplementary Software 1), which re-computes all simulations and writes the figure PDF files.

\section*{Code availability}
Custom code used to generate the results and figures is provided with this submission as \textbf{Supplementary Software 1} (\textit{generate\_figures.ipynb}).
The notebook was tested with Python 3.11.2 and the following key packages: NumPy 1.24.0, SciPy 1.14.1, NetworkX 2.8.8, and Matplotlib 3.7.5.
No restrictions apply for peer review use; for publication, the author intends to archive the code in a public repository and will update the Code availability statement accordingly.

\section*{Author contributions}
M.F.\ conceived the study, developed the theory, performed simulations, analyzed the results, and wrote the manuscript.

\section*{Competing interests}
The author declares no competing interests.

\appendix

\section{Generating-function recursion}
\label{app:gf}

Starting from \eqref{eq:Pt_recursion}, define $G_m(z)=\sum_{r\ge 0} P_m(r) z^r$.
Then
\begin{align*}
G_{m+1}(z)
&=\sum_{r\ge 0} P_m(r)\bigl(1-\mathrm{e}^{-\alpha r}\bigr) z^r \\
&\quad + \sum_{r\ge 0} P_m(r-1)\mathrm{e}^{-\alpha(r-1)} z^r\\
&=\sum_{r\ge 0} P_m(r)z^r-\sum_{r\ge 0} P_m(r)\bigl(\mathrm{e}^{-\alpha}z\bigr)^r \\
&\quad + z\sum_{r\ge 0} P_m(r)\bigl(\mathrm{e}^{-\alpha}z\bigr)^r\\
&=G_m(z)+(z-1)G_m(\mathrm{e}^{-\alpha}z),
\end{align*}
which is \eqref{eq:GF_recursion}.

\section{Exact inverse-process analysis of $\langle k\rangle$, $\Var(k)$, and $p_m$}
\label{app:moment}

We replace the mean-field closure by an exact representation of the acceptance process.
Throughout this appendix fix $\alpha>0$ and write
\[
q:=\mathrm{e}^{-\alpha}\in(0,1).
\]
Let $R_m$ denote the number of accepted edges after $m$ proposals, so that the induced
in-degree is $k=R_{N-1}$, and let
\[
G_m(z)=\sum_{r\ge 0} P_m(r)\,z^r
\]
be the probability generating function from Eq.~\eqref{eq:GF_recursion}.

\subsection{Exact derivation of Eq.~\eqref{eq:pt_accept_rank}}

Conditioning on $R_{m-1}=r$, the $m$-th proposal is accepted with probability
$q^r=\mathrm{e}^{-\alpha r}$. Averaging over $r$ gives
\[
p_m
:=\Prob(\text{$m$-th proposal is accepted})
=\sum_{r\ge 0} P_{m-1}(r)\,q^r
=G_{m-1}(q),
\]
which is exactly Eq.~\eqref{eq:pt_accept_rank}.

\subsection{Exact renewal representation}

For asymptotic analysis it is convenient to continue the same acceptance rule beyond the
first $N-1$ proposals and view $R_m$ as defined for all $m\ge 0$; the degree at system
size $N$ is then simply $k=R_{N-1}$.

Define the hitting times
\[
T_0:=0,
\qquad
T_{k+1}:=\inf\{m>T_k:R_m=k+1\},
\]
and the inter-arrival times
\[
W_k:=T_{k+1}-T_k.
\]
Once the process has accepted $k$ edges, each subsequent proposal is accepted with the
fixed probability $q^k$ until the next success. Therefore
\[
\Prob(W_k=\ell)=(1-q^k)^{\ell-1}q^k,
\qquad \ell=1,2,\dots,
\]
so $W_0,W_1,\dots$ are independent geometric random variables (with $W_0\equiv 1$).
Consequently,
\[
T_k=\sum_{r=0}^{k-1} W_r,
\qquad
R_m=\max\{k:T_k\le m\},
\]
and
\[
R_m\ge k \iff T_k\le m.
\]

The exact moments of $T_k$ are
\begin{align}
\E[T_k]
&=\sum_{r=0}^{k-1} q^{-r}
=\frac{q^{-k}-1}{q^{-1}-1}
=\frac{\mathrm{e}^{\alpha k}-1}{\mathrm{e}^{\alpha}-1},
\label{eq:ETk_exact}\\
\Var(T_k)
&=\sum_{r=0}^{k-1}\frac{1-q^r}{q^{2r}}
=\sum_{r=0}^{k-1}\bigl(\mathrm{e}^{2\alpha r}-\mathrm{e}^{\alpha r}\bigr)
=\frac{\mathrm{e}^{2\alpha k}-1}{\mathrm{e}^{2\alpha}-1}
-\frac{\mathrm{e}^{\alpha k}-1}{\mathrm{e}^{\alpha}-1}.
\label{eq:VTk_exact}
\end{align}

\subsection{Exact exponential moment identity}

Evaluating Eq.~\eqref{eq:GF_recursion} at $z=\mathrm{e}^{\alpha}=q^{-1}$ gives
\[
G_{m+1}(\mathrm{e}^{\alpha})
=
G_m(\mathrm{e}^{\alpha})+(\mathrm{e}^{\alpha}-1)G_m(1)
=
G_m(\mathrm{e}^{\alpha})+\mathrm{e}^{\alpha}-1,
\]
since $G_m(1)=1$. Because $G_0(\mathrm{e}^{\alpha})=1$, induction yields the exact identity
\begin{equation}
\E[\mathrm{e}^{\alpha R_m}]
=
G_m(\mathrm{e}^{\alpha})
=
1+(\mathrm{e}^{\alpha}-1)m.
\label{eq:exp_moment_exact}
\end{equation}

Introduce the deterministic centering
\[
x_m
:=
\frac{1}{\alpha}\log\!\bigl(1+(\mathrm{e}^{\alpha}-1)m\bigr).
\]

\subsection{Rigorous proof of Eq.~\eqref{eq:mean_scaling}}

We first derive two-sided exponential tail bounds around $x_m$.
For the upper tail, Markov's inequality and Eq.~\eqref{eq:exp_moment_exact} give, for every
$\ell\in\mathbb N_0$,
\begin{equation}
\Prob\!\bigl(R_m\ge \lceil x_m\rceil+\ell\bigr)
\le \mathrm{e}^{-\alpha\ell}.
\label{eq:upper_tail_Rm}
\end{equation}
Indeed,
\[
\Prob(R_m\ge \lceil x_m\rceil+\ell)
\le
\frac{\E[\mathrm{e}^{\alpha R_m}]}{\mathrm{e}^{\alpha(\lceil x_m\rceil+\ell)}}
\le
\mathrm{e}^{-\alpha\ell}.
\]

For the lower tail, let $k=\lfloor x_m\rfloor-\ell$; when $k<0$ the claim is trivial.
Since $R_m<k$ implies $T_k>m$, Markov's inequality and Eq.~\eqref{eq:ETk_exact} yield
\begin{equation}
\Prob\!\bigl(R_m<\lfloor x_m\rfloor-\ell\bigr)
\le \mathrm{e}^{-\alpha\ell},
\qquad \ell\in\mathbb N_0.
\label{eq:lower_tail_Rm}
\end{equation}
Indeed,
\[
\Prob(R_m<k)
=
\Prob(T_k>m)
\le
\frac{\E[T_k]}{m}
=
\frac{\mathrm{e}^{\alpha k}-1}{(\mathrm{e}^{\alpha}-1)m}
\le
\mathrm{e}^{-\alpha\ell},
\]
because $k\le x_m-\ell$.

Combining Eqs.~\eqref{eq:upper_tail_Rm} and \eqref{eq:lower_tail_Rm}, standard tail-sum
bounds imply that there exists a finite constant $C_2(\alpha)$ such that
\begin{equation}
\sup_{m\ge 0}\E[(R_m-x_m)^2]\le C_2(\alpha).
\label{eq:l2_centering_bound}
\end{equation}

Hence
\[
\bigl|\E[R_m]-x_m\bigr|
\le
\E[|R_m-x_m|]
\le
C_2(\alpha)^{1/2},
\]
so
\begin{equation}
\E[R_m]
=
\frac{1}{\alpha}\log\!\bigl(1+(\mathrm{e}^{\alpha}-1)m\bigr)+O_\alpha(1)
=
\frac{1}{\alpha}\log m+O_\alpha(1).
\label{eq:mean_asymp_exact}
\end{equation}
Evaluating at $m=N-1$ gives
\[
\langle k\rangle
=
\E[R_{N-1}]
\sim \frac{1}{\alpha}\log N,
\]
which is Eq.~\eqref{eq:mean_scaling}.

\subsection{Rigorous boundedness of the variance}

Equation~\eqref{eq:l2_centering_bound} immediately implies
\[
\Var(R_m)
\le
2\,\E[(R_m-x_m)^2]+2\,(\E[R_m]-x_m)^2
\le
4\,C_2(\alpha),
\]
uniformly in $m$. Therefore
\[
\Var(k)
=
\Var(R_{N-1})
=
O_\alpha(1)
\qquad (N\to\infty),
\]
which is Eq.~\eqref{eq:var_scaling}. Thus the exact conclusion is that the variance
remains bounded for each fixed $\alpha>0$. Numerically, the saturation value is often
close to $(2\alpha)^{-1}$ in the sparse regime, but that constant is not obtained from
an exact derivation.

\subsection{Rigorous proof of Eq.~\eqref{eq:pt_asymptotic}}

From the exact mean recursion,
\[
\E[R_{m+1}]-\E[R_m]
=
\E[q^{R_m}]
=
p_{m+1},
\]
so
\[
\E[R_m]
=
\sum_{j=1}^{m} p_j.
\]
Moreover, $p_m$ is decreasing in $m$, because $R_m$ is nondecreasing pathwise and the
map $r\mapsto q^r$ is decreasing.

By Eq.~\eqref{eq:mean_asymp_exact},
\[
\sum_{j=1}^{m} p_j
=
\E[R_m]
\sim \frac{1}{\alpha}\log m.
\]
A standard monotone density theorem for sequences\footnote{See, e.g., N.~H.~Bingham, C.~M.~Goldie, and J.~L.~Teugels, \emph{Regular Variation} (Cambridge University Press, 1989), Sec.~1.7, for equivalent formulations of the discrete monotone-density theorem / Karamata Tauberian theorem.} (equivalently, the discrete monotone
case of Karamata's Tauberian theorem) states that if $a_m$ is positive and decreasing
and $\sum_{j=1}^{m} a_j\sim c\log m$, then $a_m\sim c/m$.
Applying this to $a_m=p_m$ with $c=1/\alpha$ yields
\[
p_m\sim \frac{1}{\alpha\,m},
\qquad m\to\infty,
\]
which is Eq.~\eqref{eq:pt_asymptotic}.

\section{Computing basin (hitting) probabilities}
\label{app:committor}

Equation~\eqref{eq:committor} defines a linear system of size $(N-1)$ for the unknown vector $(u_1,\dots,u_{N-1})$.
For larger $N$ we estimate $u_a$ by Monte Carlo simulation of the Markov chain \eqref{eq:binom_chain}, which is computationally efficient because the state is one-dimensional.

\section{Two-dimensional torus embedding}
\label{app:torus2d}

To demonstrate that the spatial conclusions in Section~6 are not specific to a one-dimensional ring, we repeated the analysis on a two-dimensional $L\times L$ torus (periodic square lattice), proposing candidate sources in increasing Euclidean distance on the torus, with random tie-breaking inside each equal-distance shell.
As in the ring case, the induced in-degree distribution remains exactly $P_{\alpha}(k)$ because the crowding rule depends only on the current accepted count and is therefore invariant under permutations of the candidate list.

\begin{figure}[t]
  \centering
  \includegraphics[width=\columnwidth]{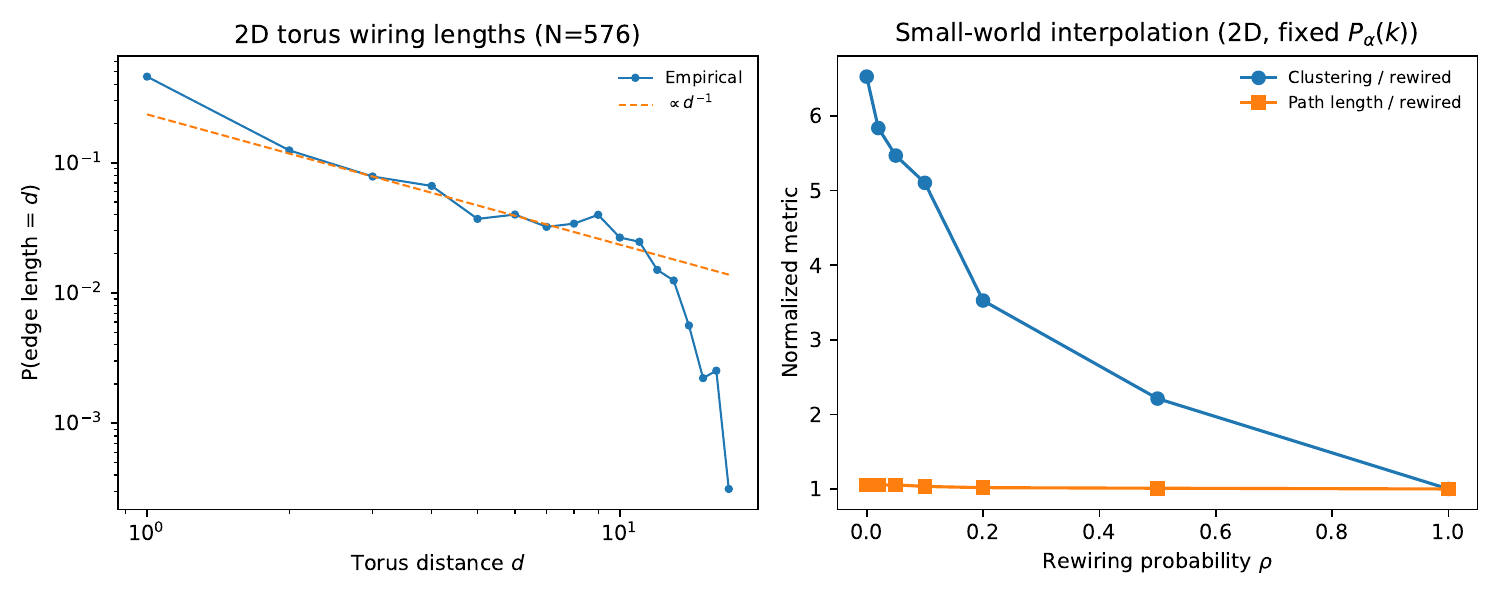}
  \caption{Two-dimensional torus embedding of the crowding architecture ($L=24$, $N=L^2=576$, $\alpha=2.66$).
  (a) Edge-length distribution under distance-ordered proposals ($\rho=0$) is well approximated by $P(d)\propto d^{-1}$.
  (b) Shortcut rewiring interpolates from the spatially ordered case to a fully rewired graph while preserving $P_{\alpha}(k)$; the plotted clustering coefficient and characteristic path length are the mean local clustering coefficient and mean shortest-path length of the symmetrized simple graph (averaged over connected node pairs), normalized by the fully rewired ($\rho=1$) baseline.}
  \label{fig:torus2d_structure}
\end{figure}

Figure~\ref{fig:torus2d_structure} confirms that the marginal wiring-length distribution remains close to $P(d)\propto 1/d$ and that a shortcut-rewiring interpolation produces a small-world regime (high clustering with short path length) while preserving $P_{\alpha}(k)$.

\section{Robustness to softened spatial ordering}
\label{app:softbias}
\begin{figure}[t]
  \centering
  \includegraphics[width=\columnwidth]{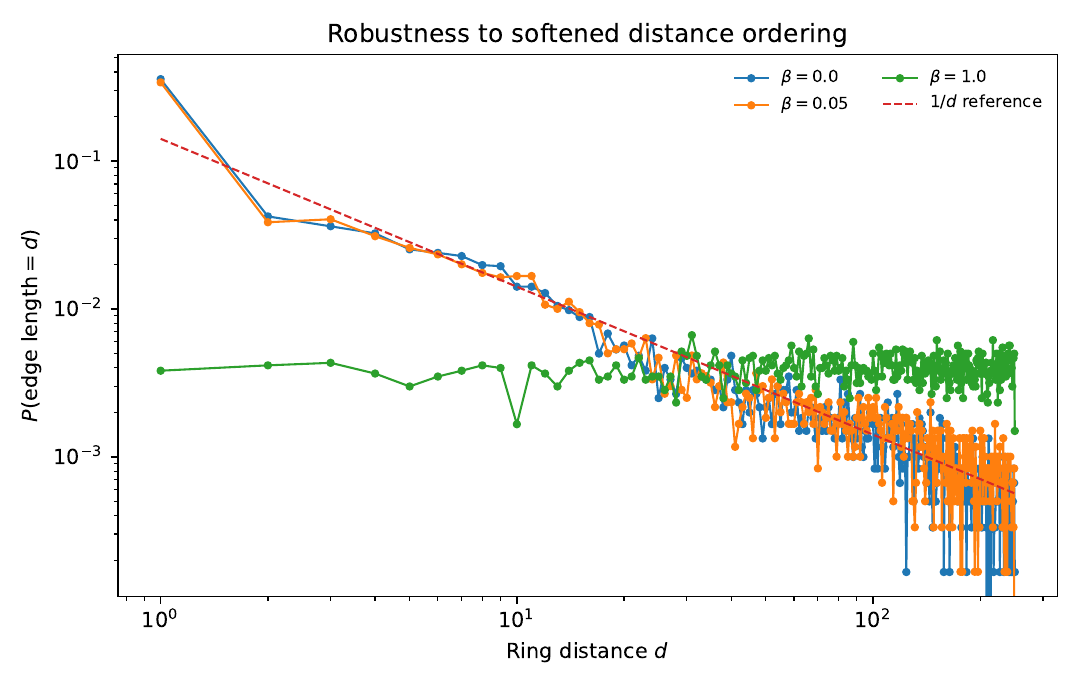}
  \caption{Robustness of the wiring-length distribution to softened distance ordering on a ring ($N=500$, $\alpha=2.66$).
  Candidate sources are proposed nearest-first with probability $1-\beta$ and uniformly at random among remaining candidates with probability $\beta$.
  For small $\beta$, the marginal wiring-length distribution remains close to $P(d)\propto 1/d$ (log-uniform lengths), while for $\beta\to 1$ it approaches the random-order baseline.
  Dashed line indicates a $1/d$ reference slope.}
  \label{fig:softbias}
\end{figure}

The nearest-first ordering used in Section~6 is an idealization.
To test robustness, we consider a softened proposal mechanism on a ring in which each proposal is drawn from the next nearest remaining candidate with probability $1-\beta$ and from a uniformly random remaining candidate with probability $\beta$.
This produces a continuum between strict distance ordering ($\beta=0$) and a random candidate order ($\beta=1$), while leaving the in-degree distribution $P_{\alpha}(k)$ unchanged because the acceptance probability depends only on the current accepted count.
Figure~\ref{fig:softbias} shows that the approximate $P(d)\propto 1/d$ length scaling persists for small but nonzero $\beta$ and gradually crosses over to the distance profile of the random-order baseline as $\beta\to 1$.

\bibliographystyle{apsrev4-2}
\bibliography{refs}

\end{document}